\documentstyle[epsfig,12pt]{article}

\def\NPB{{\em Nucl. Phys.} B~}

\def\PRL{{\em Phys. Rev. Lett.}~}

\def\PRD{{\em Phys. Rev.} D~}

\newcommand{\be}{\begin{eqnarray}}
\newcommand{\ee}{\end{eqnarray}}

\def\lsim{\mathrel{\rlap{\lower4pt\hbox{\hskip1pt$\sim$}}
    \raise1pt\hbox{$<$}}}               
\def\gsim{\mathrel{\rlap{\lower4pt\hbox{\hskip1pt$\sim$}}
    \raise1pt\hbox{$>$}}}               

\begin{document}

\rightline{{\Large Preprint RM3-TH/02-20}}

\vspace{2cm}

\begin{center}

\LARGE{A nuclear physicist on the lattice\footnote{Proceedings of the IX Italian Workshop on {\em Problems in Theoretical Nuclear Physics}, Cortona (Italia), October 9-12, 2002, World Scientific Publishing (Singapore), in press.}}

\vspace{1cm}

\large{Silvano Simula}\\

\vspace{0.5cm}

\normalsize{INFN, Sezione Roma III, Via della Vasca Navale 84, I-00146 Roma, Italy}

\vspace{1cm}

\normalsize{for the $SPQ_{cd}R$ collaboration} 

\vspace{1cm}

\end{center}

\vspace{1cm}

\begin{abstract}

\noindent Nuclear and lattice physicists share several topics of common interest, like hadronic masses, electroweak form factors and structure functions of hadrons. The main physics issues that a new collaboration between a nuclear physicist and the lattice group of the $SPQ_{cd}R$ collaboration is going to address, are summarized and preliminary results obtained for meson and baryon masses are presented. The importance of adopting non-perturbative approaches based on the fundamental theory of the strong interaction, like lattice formulations of $QCD$, is stressed.

\end{abstract}

\newpage

\pagestyle{plain}

\section{From nuclei to quarks and gluons}

\indent Since several years an increasing part of our nuclear physics community is deeply involved in the study of several topics pertaining to hadronic physics, like hadron mass spectra and decays, form factors of the nucleon and its resonances, structure functions of hadrons, quark-gluon plasma, etc. Since Quantum ChromoDynamics ($QCD$) is believed to be our fundamental theory of the strong interaction, quarks and gluons are the degrees of freedom to be used in the description of the hadron structure. This fact implies two requirements. The first one is to describe the hadron structure through a non-perturbative approach based on the fundamental theory. The second one is that the same approach should be used to calculate different matrix elements relevant for $QCD$. Both requirements are fulfilled by lattice formulations of $QCD$. Thus it is clear that {\em lattice $QCD$ represents the future of theoretical nuclear physics}.

\indent The aim of this contribution is to point out that there are several important physics issues shared by the nuclear and lattice communities. In this respect a new collaboration has recently started between myself and the lattice group of the Southampton-Paris-QCD-Rome ($SPQ_{cd}R$) collaboration with the aim of addressing: ~ i) the meson and baryon mass spectra; ~ ii) the electroweak form factors and structure functions of the baryon octet and decuplet; ~ iii) the electric dipole moment of the neutron induced by $CP$ violations both in the electroweak and in the strong sectors (see in the latter case Ref.~\cite{dipole}).

\indent After recalling the main features of the Wilson and Clover formulations of $QCD$ on the lattice our preliminary results obtained for meson and baryon masses in the quenched approximation will be presented.
 
\section{Lattice $QCD$}

\indent Lattice formulations represent a possible way to regularize the ultraviolet divergencies of a quantum field theory. At the same time they are a very powerful tool to investigate nonlinear field theories in their own non-perturbative domain. As far as $QCD$ is concerned, various lattice formulations have been proposed in the literature and are under continuous investigation and improvement. The $SPQ_{cd}R$ collaboration adopts the Wilson formulation of the $QCD$ action \cite{Wilson}, in which the total action is given by the sum of a fermionic and a gluonic part
 \be
      S_{Wilson} = S_{\psi} + S_G ~ ,
      \label{eq:action}
 \ee
which in Euclidean space read as
 \be
      S_{\psi} & = & a^4 \sum_{x, f} \left\{ -{1 \over 2a} \sum_{\mu =1}^{4} 
      \left[ \bar{\psi}_f(x) (r - \gamma_{\mu}) U_{\mu}(x) \psi_f(x + 
      \hat{\mu}) \right. \right. \nonumber \\
      & + & \left. \left. \bar{\psi}_f(x + \hat{\mu}) (r + \gamma_{\mu}) 
      U_{\mu}^{\dagger}(x) \psi_f(x) \right] + \bar{\psi}_f(x) \left[ m_f^0 
      + {4r \over a} \right] \psi_f(x) \right\}
      \label{eq:fermionic}
 \ee
and
 \be
      S_G = {6 \over g_0^2} \sum_{plaquette} \left\{ 1 - {1 \over 3} 
      \mbox{Re} \left[ \mbox{Tr} ~ U_{plaquette} \right] \right\} ~ .
      \label{eq:gluonic}
 \ee
where $\psi_f$ is the fermionic field with flavor $f$, $m_f^0$ is the bare quark mass, $g_0$ is the bare coupling constant, $a$ is the lattice spacing and $U_{\mu}(x) \equiv exp(i g_0 a A_{\mu}^c \lambda^c / 2)$ is the gauge link. In Eq. (\ref{eq:fermionic}) the terms proportional to the parameter $r$ represent the so-called Wilson term, which avoids the fermion doubling problem. The Wilson term vanishes in the continuum limit ($a \to 0$), but breaks chiral symmetry. In Eq. (\ref{eq:gluonic}) the sum extends over the symmetric plaquettes, i.e. all the plaquettes having $x$ as one corner with the plaquette operator given by the product of the four gauge links comprising the plaquette in the clockwise sense.

\indent In the continuum limit both $S_{\psi}$ and $S_G$ reduce themselves to the fermionic and gluonic part of the $QCD$ action, respectively. However, the convergence is different for $S_{\psi}$ and $S_G$, namely
 \be
      S_{\psi} & \to & \left[ S_{QCD} \right]_{\psi} + O(a) ~ , \nonumber \\
      S_G & \to & \left[ S_{QCD} \right]_G + O(a^2) ~ .
      \label{eq:limits}
 \ee
Therefore, a modification of the Wilson action has been proposed in order to kill all the linear terms in the lattice spacing. The new action is known as the Clover action \cite{Clover} and reads explicitly as
 \be
      S_{Clover} = S_{Wilson} - c_{SW} a^4 \sum_{x, f} i g_0 {a r \over 4} 
      \sum_{\mu \nu} \bar{\psi}_f(x) \sigma_{\mu \nu} P_{\mu \nu} \psi_f(x)
     \label{eq:clover}
 \ee
where $P_{\mu \nu}$ is the plaquette operator and $c_{SW}$ is a parameter which depends on $g_0$ and should be determined non-perturbatively. This was done in Ref. \cite{CSW} and one finally  gets
 \be
      S_{Clover} \to S_{QCD} + O(a^2)
      \label{eq:clover_limit}
 \ee

\indent Thus, the Clover action (\ref{eq:clover}) makes the convergence to the continuum limit faster as fas as the action is concerned. However, the improvement provided by the Clover action does not exhaust the full improvement program, because in general fermionic operators have to be improved themselves in order to guarantee that their matrix elements are not affected by linear terms in the lattice spacing. Nevertheless, as far as hadronic masses are concerned, it is enough to consider only the improvement of the action.

\indent Hadronic masses can be determined by analyzing the time evolution of the two-point function at zero-momentum
 \be
      G_H(t) = \sum_{\vec{x}} \langle 0 | \hat{T} \{ O_H(\vec{x}, t) ~ 
      \overline{O}_H(\vec{0}, 0) \} | 0 \rangle
      \label{eq:2point}
 \ee
where $\hat{T}$ stands for the time-ordered product and $O_H(\vec{x}, t)$ is an interpolating field having the same quantum numbers of the hadron $H$. Inserting a complete set of hadronic states one gets $G_H(t) = \sum_n Z_n^H exp(-M_n^H t)$, where $M_n^H$ is the mass of the $n$-th eigenstate with the quantum numbers of the hadron $H$ and $Z_n^H$ is the overlap of the interpolating field with the $n$-th eigenstate. Therefore, for large source times the ground state is isolated, i.e. $G_H(t) \to_{t >> a} Z_0^H exp(-M_0^H t)$. The value $M_0^H$ can be obtained from the {\em plateau} of the effective mass $M_{eff}(t)$, defined as
 \be
      M_{eff}(t/a) = \mbox{ln}[ G(t - a) / G(t) ] \to_{t >> a} M_0^H \cdot a
      \label{eq:effective}
 \ee

\section{Meson masses}

\indent We have carried out $QCD$ simulations in the quenched approximation using $320$ gauge configurations in a $24^3 \times 64$ lattice volume  with a bare coupling constant $g_0$ corresponding to $\beta \equiv 6 / g_0^2 = 6.0$.  We have considered various values of the hopping parameter $K$, which is connected to the (lattice) bare quark mass $m$ by
 \be
       m = {1 \over 2a} \left[ {1 \over K} - {1 \over K_c} \right]
       \label{eq:hopping}
 \ee
where $K_c$ is the critical value of the hopping parameter corresponding to the chiral point $m = 0$\footnote{The lattice mass $m$ differs from the bare mass $m^0$ of Eq.~(\ref{eq:fermionic}). Indeed, since the Wilson term breaks chiral symmetry, the quark mass receives an additive renormalization, i.e. $m = m^0  - \overline{M}$, where $\overline{M}$ diverges in the continuum limit. Thus, using the Wilson (or Clover) formulation the chiral limit is not given by $m^0 \to 0$, but instead by $m^0 = \overline{M}$.}. In what follows $SU(3)$-flavor symmetry is assumed.

\indent In case of pseudoscalar ($PS$) and vector ($V$) mesons we have used the standard interpolating fields
 \be
      O_{PS}(\vec{x}, t) & = & \bar{u}(x) \gamma_5 d(x) ~ , \nonumber \\
      O_V(\vec{x}, t) & = & \bar{u}(x) \gamma_{\mu} d(x) ~ .
      \label{eq:Omesons}
 \ee
Taking into account the periodicity of the lattice the two-point function has the following large-time behavior [$t >> a$ and $(T - t) >> a$]
 \be
      G(t) \to Z_0 \left[ e^{-M_0 t} + e^{-M_0 (T - t)} \right]
      \label{eq:Gmesons}
 \ee
where in our case $T = 64 \cdot a$. From Eq.~(\ref{eq:Gmesons}) one can see that the first and second halves of the lattice time contain the same information. The effective mass obtained for $PS$ and $V$ mesons at $K = 0.13300$ is reported in Fig.~1 to illustrate the quality of the plateau. 

\begin{figure}[t]

\centerline{\epsfxsize=16cm \epsfbox{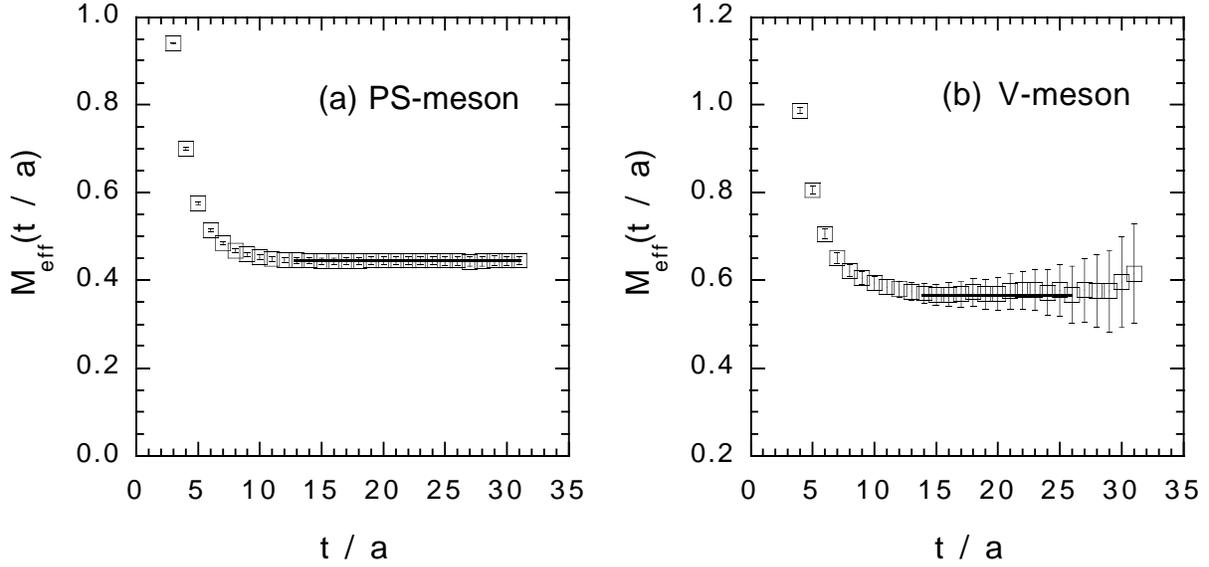}}

\caption{Effective mass for $PS$ (a) and $V$ (b) mesons versus the source time $t$ in units of the lattice spacing $a$ for $K = 0.13300$ . The solid lines show the part of the plateau used to extract the ground-state masses.}

\end{figure}

\indent The values obtained with the plateau method for the $PS$- and $V$-meson masses are reported in Fig.~2, where it can be seen that our results are in very good agreement with those obtained in Refs.~\cite{DESY,UKQCD} at the same value of $\beta$. The squared $PS$ masses can be nicely fitted by a linear relation with the inverse hopping parameter [see Fig.~2(a)]. This means that $M_{PS}^2$ is almost proportional to the lattice bare mass $m$. The intercept of the linear slope shown in Fig.~2(a) corresponds to $K = K_c = 0.13526 ~ (2)$. From Fig.~2(b) it can be seen that the calculated $V$-meson masses exhibit a linear relation with the squared $PS$-meson mass, viz.
 \be
      M_V \cdot a = A + B \left( M_{PS} \cdot a \right)^2
      \label{eq:vector}
 \ee
where $A$ and $B$ are two coefficients fitted to our lattice results. Then, using the physical masses of $K$ and $K^*$ mesons we obtain for the inverse lattice spacing the value $a^{-1} = 1.90 \pm 0.08 ~ GeV$, which corresponds to $a = 0.104 \pm 0.004 ~ fm$. Moreover, using the physical masses of the pion and the kaon the lattice bare masses can be determined from the slope shown in Fig.~2(a), obtaining $m_l \equiv (m_u + m_d) /2 = 3.3 \pm 0.1 ~ MeV$ and $m_s = 78 \pm 3 ~ MeV$. These masses have still to be renormalized in some scheme. Adopting the $\overline{MS}$ scheme one has $m^{\overline{MS}}(\mu) = Z^{\overline{MS}}(a \mu) \cdot m$, where $\mu$ is the renormalization scale (conventionally taken at the value $\mu = 2 ~ GeV$) and $Z^{\overline{MS}}$ is a renormalization coefficient which has been calculated nonperturbatively in Ref.~\cite{Vittorio}. At $\beta = 6.0$ we get $m_n^{\overline{MS}}(2 ~ GeV) = 4.1 \pm 0.1 ~ MeV$ and $m_s^{\overline{MS}}(2 ~ GeV) = 96 \pm 4 ~ MeV$ which compare favorably with world averages (cf. Ref.~\cite{Vittorio}).

\begin{figure}[t]

\centerline{\epsfxsize=16cm \epsfbox{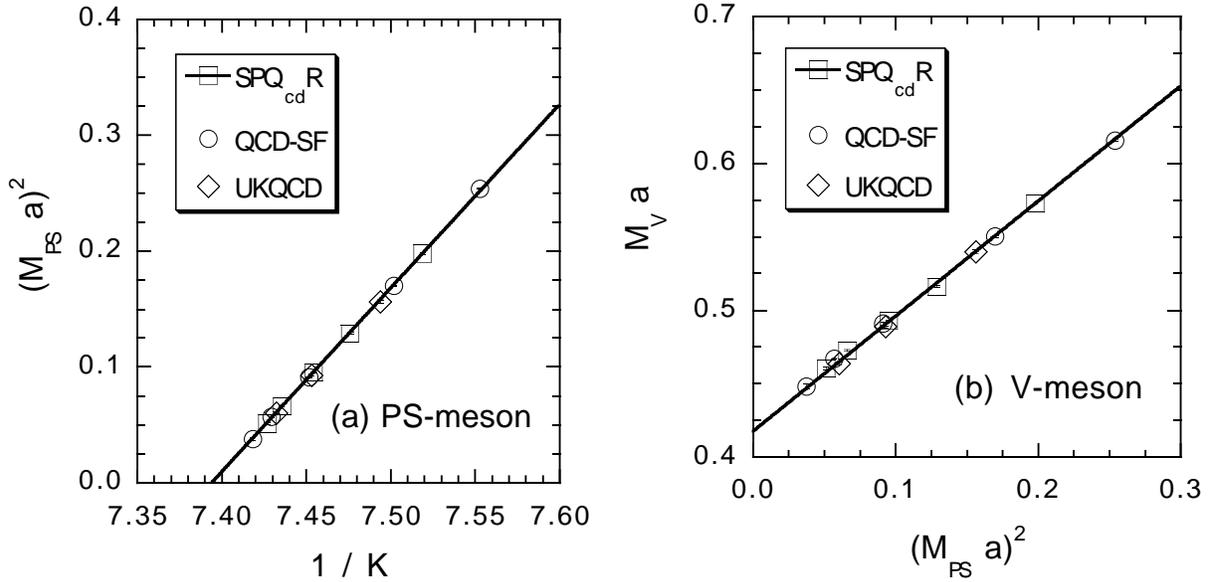}}

\caption{(a) Squared $PS$-meson mass versus the inverse hopping parameter; (b) $V$-meson mass versus the squared $PS$ mass in units of the lattice spacing $a$ . The solid lines are a linear fit to our results (squares). Dots and diamonds correspond to the results of Refs.~\protect\cite{DESY,UKQCD}, respectively.}

\end{figure}

\section{Baryon masses}

\indent In case of baryons we have considered two interpolating fields, having the same quantum numbers of the proton and of its negative-parity partner (the $S_{11}$ resonance), namely:
 \be
      O_{N^+}(\vec{x}, t) & = & {1 \over \sqrt{6}} \varepsilon_{a b c} 
      \left[ u_a^T C \gamma_5 d_b \right] u_c ~ , \nonumber \\
      O_{N^-}(\vec{x}, t) & = & {1 \over \sqrt{6}} \varepsilon_{a b c} 
      \left[ u_a^T C \gamma_5 d_b \right] \gamma_5 u_c ~ ,
      \label{eq:baryons}
 \ee
where $C$ is the charge conjugation matrix. However, one has $O_{N^+}(\vec{x}, t) = - \gamma_5 O_{N^-}(\vec{x}, t) \gamma_5$, which implies a structure of the baryon two-point function different from the meson one [see Eq.~(\ref{eq:Gmesons})]. Indeed, for $t >> a$ and $(T - t) >> a$ one has
 \be
      G(t) & \to & {1 + \gamma_0 \over 2} \left[ Z_+ e^{-M^+ t} - Z_- 
      e^{-M^- (T - t)} \right] \nonumber \\
      & + & {1 - \gamma_0 \over 2} \left[ Z_+ e^{-M^+ (T - t)} - Z_- e^{-M^- 
      t} \right]
      \label{eq:Gbaryons}
 \ee
Thus, in the upper components the proton propagates forward while its negative partner goes backward. The situation is simply reversed in the lower components of the two-point function. Therefore, from the large-time behavior of the two-point function one can obtain simultaneously the ground-state masses of a hadron and of its negative-parity partner. Our results are collected in Fig.~3 and compared (favorably) with those of Refs.~\cite{DESY,UKQCD}. Adopting a linear fit with the squared $PS$ mass [as suggested by Fig.~3(a)], we obtain for the mass of the proton the value $0.98 \pm 0.05 ~ GeV$ and for the mass ratio $M_{N^-} / M_{N^+}$ the value $1.65 \pm 0.06$. Both results compare very favorably with the experimental data $M_p = 0.938 ~ GeV$ and $M_{S_{11}} / M_p = 1.63$ from \cite{PDG}.

\begin{figure}[t]

\centerline{\epsfxsize=16cm \epsfbox{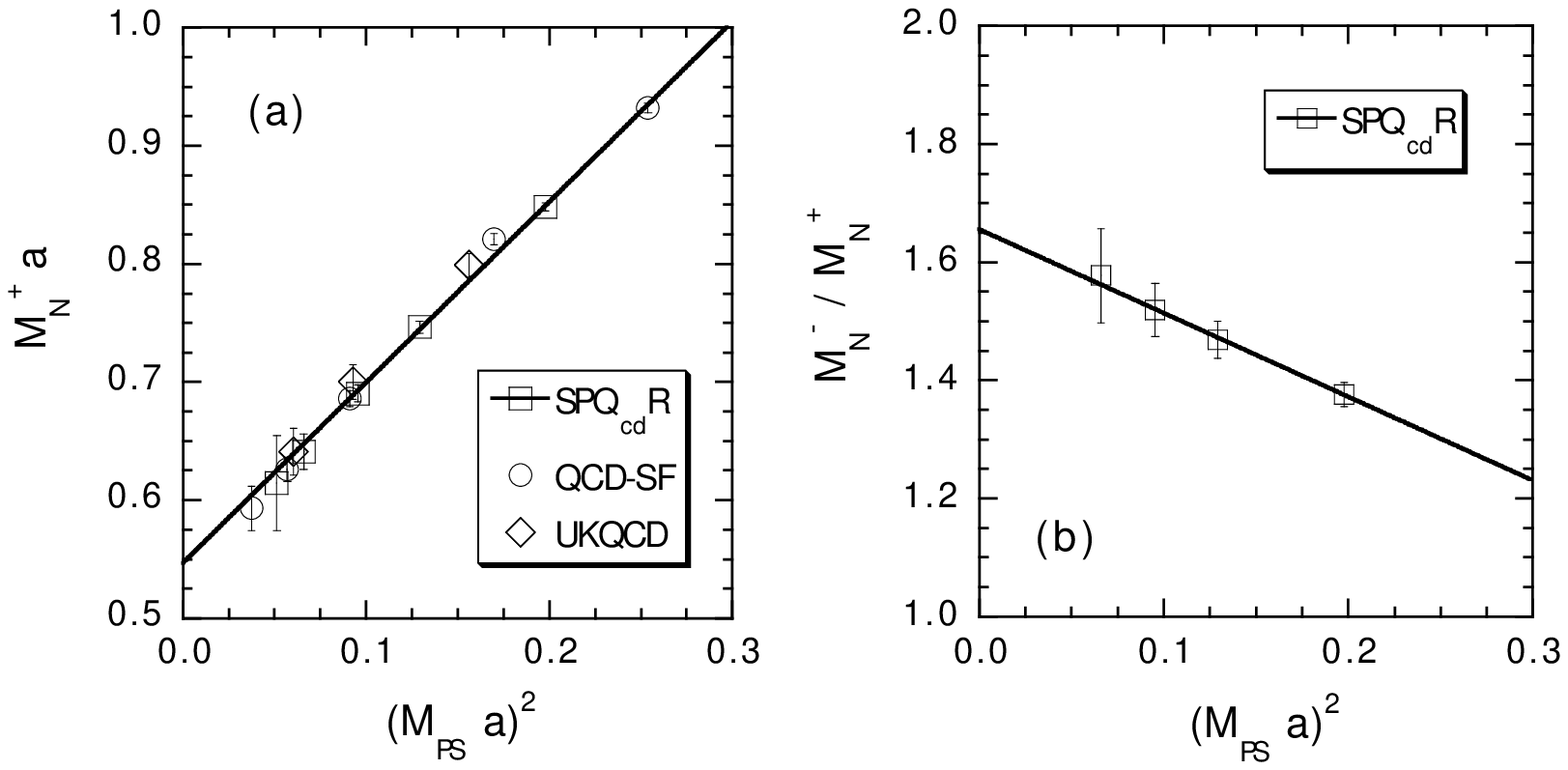}}

\caption{(a) Positive-parity nucleonic mass versus the squared $PS$ mass; (b) ratio of negative to positive parity nucleonic masses in units of the lattice spacing $a$ . The solid lines are a linear fit to our results (squares). In (a) Dots and diamonds correspond to the results of Refs.~\protect\cite{DESY,UKQCD}, respectively.}

\end{figure}

\indent In Ref.~\cite{UKQCD} the effects of $SU(3)$-flavor breaking on light-hadron masses have been investigated in details and it was shown that to a very good approximation the light-hadron masses depend linearly on the sum of the lattice bare quark masses. Therefore, using the slopes shown in Fig.~3 we can estimate the masses of strange hadrons, like the cascade $\Xi$ and the $\Lambda$ baryon. Our results for various hadron masses are collected in Fig.~4. Note that: ~ i)  the interpolating fields (\ref{eq:baryons}) do not distinguish between $\Lambda$ and $\Sigma$ baryons, and ~ ii) the masses of $\pi$, $K$, $\rho$ and $K^*$ mesons are not reported, because they are used to fix the chiral point, the lattice spacing and the lattice bare quark masses. From Fig.~4 it can be seen that several experimental masses are nicely reproduced by our quenched calculations with the only remarkable exception of the $\Lambda(1405)$ resonance. The anomalous light mass of the latter suggest a possible interpretation as a $N K$ molecule, which is totally missed in a quenched calculation.

\begin{figure}[t]

\centerline{\epsfxsize=16cm \epsfbox{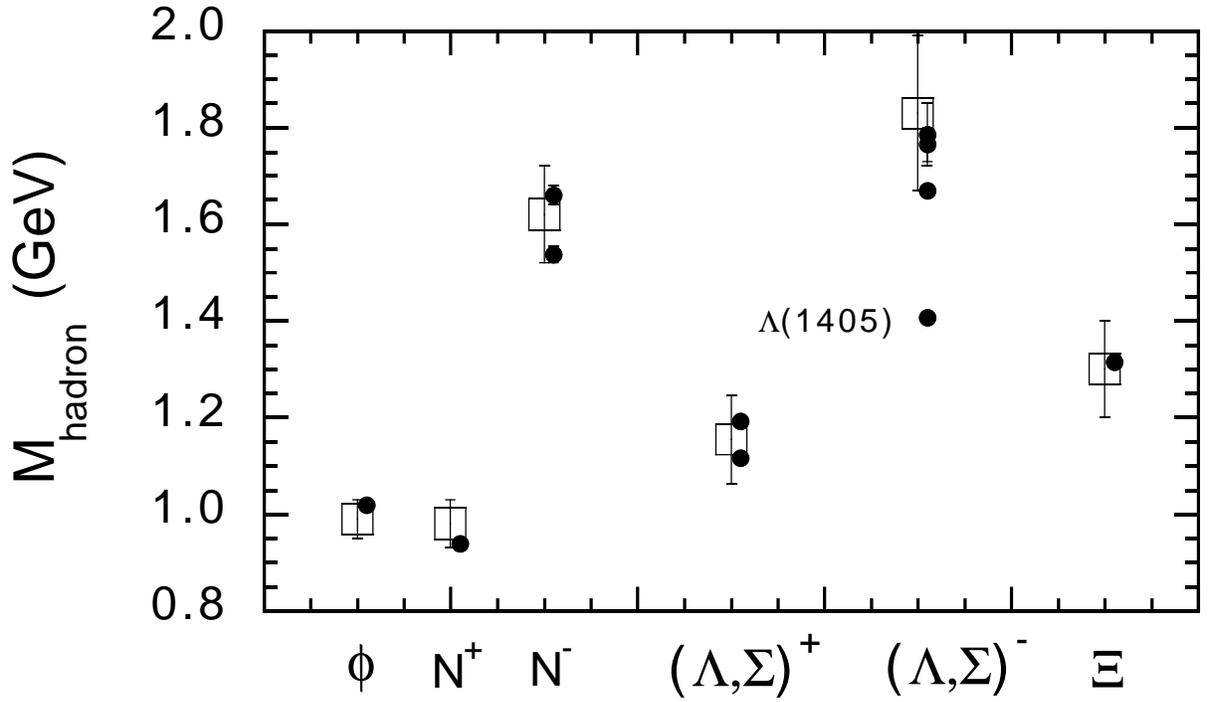}}

\caption{Mass of various light hadrons. Open squares are our lattice results, while full dots represent experimental data from $PDG$ \protect\cite{PDG}.}

\end{figure}

\indent The results presented have been obtained at $\beta = 6.0$ only. Therefore, $QCD$ simulations at larger values of $\beta$ (which correspond to lower values of the lattice spacing) are needed to assess the uncertainties related to the continuum extrapolation. We stress that thanks to the use of the Clover action (\ref{eq:clover}) the above-mentioned uncertainties are only of the order of $O(a^2)$.

\section{Conclusions}

\indent Several important physics issues are shared by nuclear and lattice communities. In this respect a new collaboration between a nuclear physicist and the lattice group of the $SPQ_{cd}R$ collaboration has recently started with the aim of addressing: ~ i) the meson and baryon mass spectra; ~ ii) the electroweak form factors and structure functions of the baryon octet and decuplet; ~ iii) the electric dipole moment of the neutron induced by $CP$ violations both in the electroweak and in the strong sectors (see for the latter case Ref.~\cite{dipole}). Preliminary results obtained for meson and baryon masses in the quenched approximation have been presented.

\indent Our main conclusion is that {\em the future of theoretical nuclear physics is represented by lattice $QCD$}, to which efforts and manpower should be devoted by our nuclear community.


\begin{thebibliography}{99}

\bibitem{dipole} D. Guadagnoli {\em et al.}, hep-lat/0210044.

\bibitem{Wilson} K.G. Wilson, in {\em New Phenomena in Subnuclear Physics}, edited by A. Zichichi, Plenum Press, New York (1977).


\bibitem{Clover} B. Sheikholeslami and R. Wohlert, \NPB {\bf 259}, 572 (1985).

\bibitem{CSW} M. L\"uscher {\em et al.}, \NPB {\bf 491}, 323 (1997). R.G. Edwards {\em et al.}, \PRL {\bf 80}, 3448 (1998). K. Jansen and R. Sommer, \NPB {\bf 530}, 185 (1998).

\bibitem{DESY} M. G\"ockeler {\em et al.}, \PRD {\bf 57}, 5562 (1998).

\bibitem{UKQCD} K.C. Bowler {\em et al.}, \PRD {\bf 62}, 054506 (2000).

\bibitem{PDG} $PDG ~ '02$: K. Hagiwara {\em et al.}, \PRD {\bf 66}, 010001 (2002).

\bibitem{Vittorio} D. Becirevic {\em et al.}, hep-lat/0208003.

\end{thebibliography}
\end{document}